\documentclass[11pt,twoside]{article}

\usepackage{asp2021}

\aspSuppressVolSlug
\resetcounters

\begin{document}

\title{An Interactive Metrics Dashboard for the Keck Observatory Archive}


\author{G.~Bruce~Berriman,$^1$ and Min Phone Myat Zaw,$^2$}
\affil{$^1$Caltech/IPAC-NExScI,~Pasadena,~CA~~91125,~USA;}~~~~~~~\email{gbb@ipac.caltech.edu}

\affil{$^2$~University of California, ~405 Hilgard Avenue, ~Los Angeles, ~CA~~ 90095, USA}


\paperauthor{G.~B.~Berriman}{gbb@ipac.caltech.edu}{0000-0001-8388-534X}{Caltech}{IPAC/NExScI}{Pasadena}{CA}{USA}

\paperauthor{M.~P.~Z.~Zaw}{myatzawowl@gmail.com}{ 0009-0005-7143-4783}{University of California}{405 Hilgard Avenue}{Los Angeles}{CA}{USA}



\begin{abstract}
Since 2004, the Keck Observatory Archive (KOA) has operated as a NASA-funded collaboration between the NASA Exoplanet Science Institute ( NExScI) and the W.M. Keck Observatory. It ingests and serves all data acquired by the twin 10-meter Keck telescopes on Mauna Kea, Hawaii. In the past three years, KOA has begun a modernization program to replace the architecture and systems used since the archive's creation with a new modern Python-based infrastructure. This infrastructure will position KOA to respond to the rapid growth of new and complex data sets that will be acquired by new instruments now in development, and enable follow-up to identify the deluge of alerts of transient sources expected by new survey telescopes such as the Vera C. Rubin Observatory. Since 2022, KOA has ingested new data in near-real time, generally within one minute of creation, and has made them immediately accessible to observers through a dedicated web interface. The archive is now deploying a new, scalable, Python-based, VO-compliant query infrastructure built with the Plotly-Dash framework and R-tree indices to speed-up queries by a factor of 20.

The project described here exploits the new query infrastructure to develop a dashboard that will return live metrics on the performance and growth of the archive. These metrics assess the current health of the archive and guide planning future hardware and software upgrades. This single dashboard will enable, for example, monitoring of real-time ingestion, as well as studying the long-term growth of the archive. Current methods of gathering metrics that have been in place since the archive opened will not support the archive as it continues to scale. These methods suffer from high latency, are not optimized for on-demand metrics, are scattered among various tools, and are cumbersome to use.

\end{abstract}



\section{Why Build A Metrics Dashboard?}

The Keck Observatory Archive (KOA) \footnote{\url {https://koa/ipac.caltech.edu}} has ingested all data acquired at the W. M. Keck Observatory \footnote{\url {https://keckobservatory.org/}}  since 2004. These data become public when the period of exclusive use by the observer expires (18 months; 12 months for NASA data).

Operations are evolving rapidly to position the archive to support new science directions in astronomy. Data are now ingested in near-real time, generally within 10 seconds of acquisition, rather than at the end of the night. The current set of metrics-gathering tools, largely based on PHP, require an administrator to operate, create only static plots, and are generally slow. They are inadequate for gathering metrics on archive content and performance in this dynamic operations environment. This paper reports on the development of a web-based dashboard that updates metrics in near real-time every 5-7 seconds. The development is part of a long-term project to modernize the archive architecture from a C-based system to one based on a Python framework.

\section{The Dashboard in Operation}

The first version of the dashboard  deploys  the architecture and returns the number of files in the archive over time.  A video demonstration of the dashboard has been posted online \footnote{\url {https://youtu.be/QlJ-67CJNfw}}. Figure \ref{ex_Fig1} is a screenshot from this video to show the important functions of the dashboard:

\articlefigure[width=0.75\textwidth]{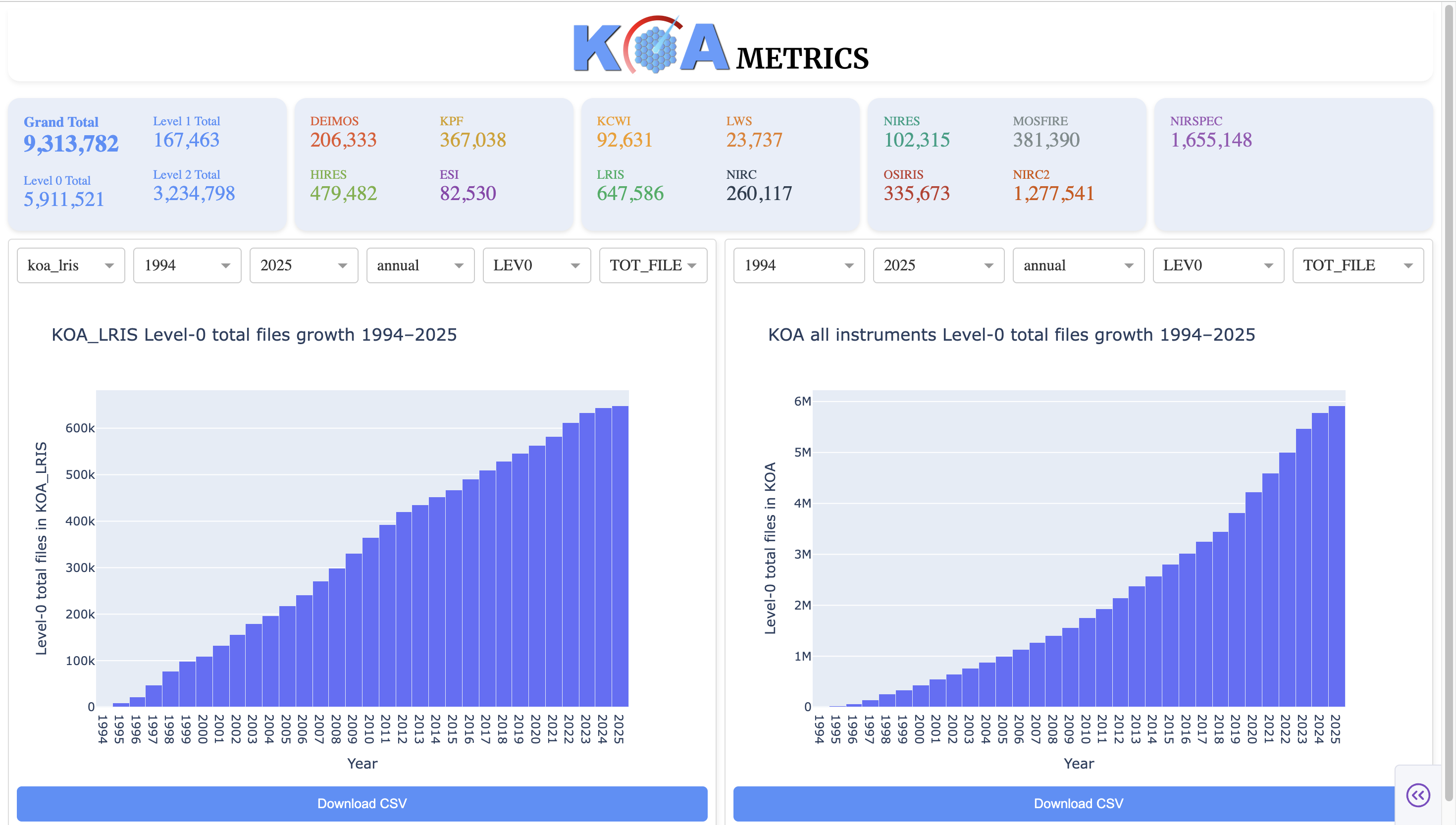}{ex_Fig1}{Screenshot of the KOA metrics dashboard showing the total number of level 0 files per instrument and plots of the number of raw files for the LRIS instrument, with parameters selected from the pull-down menu.}

\begin{itemize}
\item  The data loading spinner (top right)displays when the dashboard is being updated. 
\item Cards showing the number of level 0 files ingested per instrument and the grand total, updated every seven seconds. 
\item Pull-down menus to select  data for presentation: instrument, data type, date range and cadence (weekly, monthly), and year range for grand totals.
\item Histograms of data in the plots, available for download in CSV format. 
\end{itemize}

The remainder of this paper describes the architecture and optimization of the dashboard.

\section {Architectural Choices}

The dashboard architecture was designed with an emphasis on widely used and maintainable Open Source technologies: 

\begin{itemize} 

\item 
The dashboard is integrated  into the Plotly-Dash framework\footnote{\url{https://dash.plotly.com/}},and is the basis for the new KOA software infrastructure \citep{P412_adassxxxiv}. 
\item 
The dashboard operates with callbacks that use the Jinja 2 template engine\footnote{\url{https://jinja.palletsprojects.com/en/stable/}} to build dynamic queries.
\item 
The queries are submitted through the VO-compliant nexsciTAP server\footnote{\url{https://github.com/Caltech-IPAC/nexsciTAP}} to \\
SQLAlchemy\footnote{\url{https://www.sqlalchemy.org/}}, an SQL toolkit for Python that simplifies database tasks.
\item 
Color-coded Key Performance Indicator (KPI) cards \footnote{\url{https://www.kpi.org/}}summarize  archive contents.
\item A backend daemon polls KOA through nexsciTAP for new metrics in near real-time and ingests them upon receipt, and records these ingestions in a log.                                                                        

\end{itemize}

\section{Optimizing Performance for Updates at a Seven-Second Cadence}

A key part of the design is to the dashboard is able to support the cadence at which data are ingested. 

\subsection{Database Design}
Database tables have been created for each year of  observatory  operations. The schema has been designed to minimize data transfer  from the DBMS to the dashboard:
\begin{itemize}
    \item The time and instrument values are the primary columns.
    \item The numbers of science and calibration files are composed as virtual columns.
    \item\ SQL templates create simplified queries.
\end{itemize}



\subsection{Parallel Queries in the Backend Systems Architecture}

Pre-calculated, static database tables from 1994-2024 have been created on a daily, monthly, and annual basis. The tables for the current year, 2025 at the time of the presentation, are dynamic. Parallel queries to these two sets of tables improve performance, as shown in Figure \ref{ex_Fig2}.

\articlefigure[width=0.85\textwidth]{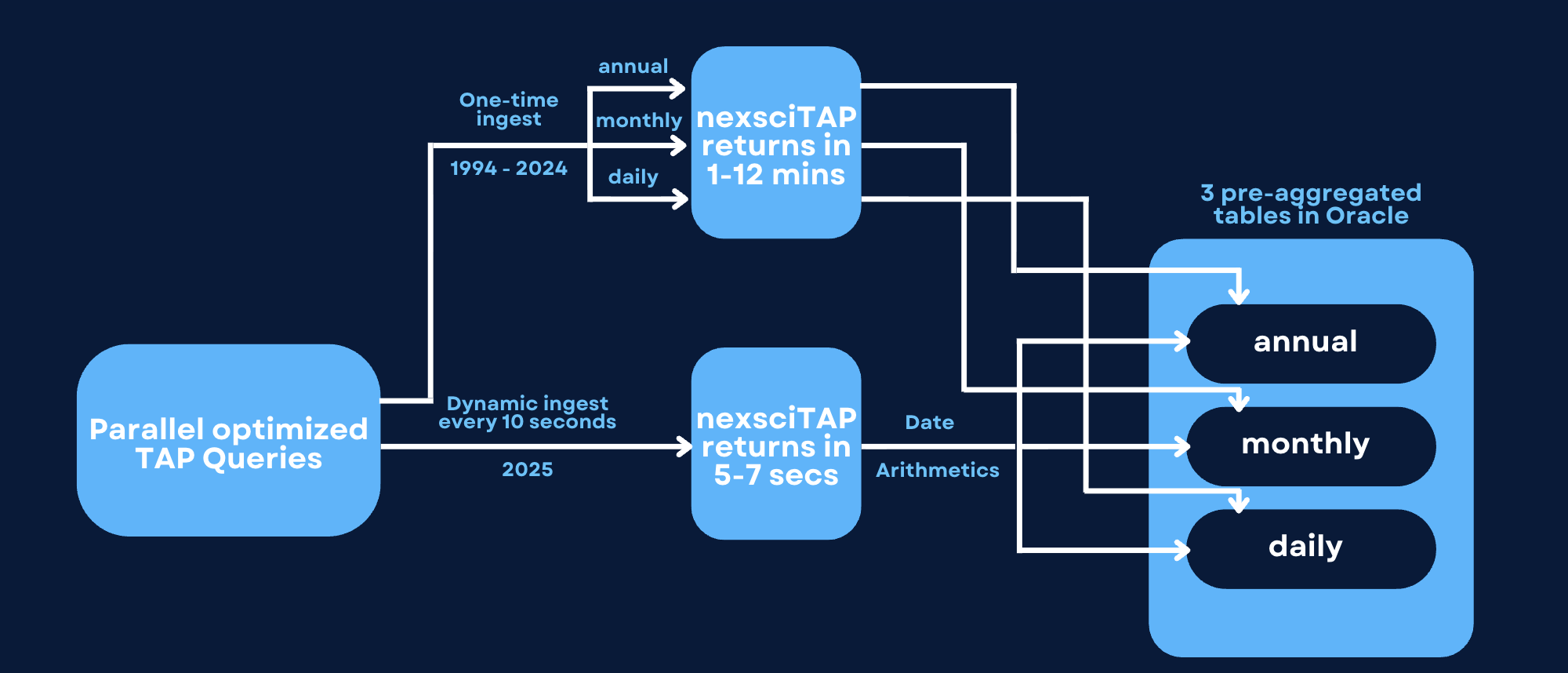}{ex_Fig2}{Block diagram of the use of parallel queries to the database that improve performance.}

Performance is continuously monitored by the ingestion daemon, and by the Plotly built-in event monitoring system, which graphically displays the update speeds in milliseconds.

\section{Conclusions}

The dashboard architecture presented here meets the archive's goal of monitoring the number of files ingested by the archive during a calendar year in near real-time. In the coming year, we will expand the dashboard to include a comprehensive set of metrics, such as data volume, the number of archive queries, and download volumes. We will also augment functionality as needed, such as overlay plots for selected instruments.

\acknowledgements The Keck Observatory Archive (KOA) is a collaboration between the NASA Exoplanet Science Institute (NExScI) and the W. M. Keck Observatory (WMKO). NExScI is sponsored by NASA's Exoplanet Exploration Program and operated by the California Institute of Technology in coordination with the Jet Propulsion Laboratory (JPL). Mr. Zaw developed this dashboard as a 2025 Caltech Summer Undergraduate Research Fellow (SURF). We thank the KOA team for supporting this project, and we thank Ms. Marcy Harbut for proof reading the paper.

The observatory was made possible by the generous financial support of the W. M. Keck Foundation. The authors wish to recognize and acknowledge the very significant cultural role and reverence that the summit of Mauna Kea has always had within the indigenous Hawaiian community. We are most fortunate to have the opportunity to conduct observations from this mountain.

\bibliography{73}  

@INPROCEEDINGS{P412_adassxxxiv,
   author     = {{Moseley}, R. and {Berriman}, G. B. and {Gelino}, C. R. and {Good}, J. C. and {Oluyide}, T.},
booktitle     = {ADASS XXXIV},
     year     = 2024,
   editor     = {{DeMarco}, A. and {Said}, J.},
   volume     = {TBD},
   series     = {ASP Conf. Ser.},
    pages     = {999 TBD},
    publisher = "ASP",
    address   = "San Francisco",
}

\end{document}